\documentclass[prl,twocolumn,amsmath,amssymb,superscriptaddress,showpacs]{revtex4-1}
\usepackage{graphicx}
\usepackage{ulem}
\usepackage{graphicx}
\usepackage{dcolumn}
\usepackage{bm}
\usepackage{mathrsfs}
\usepackage{subfigure}
\usepackage{natbib}
\usepackage{color}
\usepackage{amssymb}
\usepackage{amsmath}
\usepackage{bbm}

\begin{document}

\title{The BCS-BEC crossover induced by a shallow band: Pushing standard
superconductivity types apart}

\author{S. Wolf}
\affiliation{Institut f\"{u}r Theoretische Physik III, Bayreuth
Universit\"{a}t, Bayreuth 95440, Germany}

\author{A. Vagov}
\affiliation{Institut f\"{u}r Theoretische Physik III, Bayreuth
Universit\"{a}t, Bayreuth 95440, Germany}

\author{A. A. Shanenko}
\affiliation{Departamento de F\'isica, Universidade Federal de
Pernambuco, Av. Jorn. An\'ibal Fernandes, s/n, Cidade
Universit\'aria 50740-560, Recife, PE, Brazil}

\author{V. M. Axt}
\affiliation{Institut f\"{u}r Theoretische Physik III, Bayreuth
Universit\"{a}t, Bayreuth 95440, Germany}

\author{A. Perali}
\affiliation{School of Pharmacy, Physics Unit, University of
Camerino, I-62032-Camerino, Italy}

\author{J. Albino Aguiar}
\affiliation{Departamento de F\'isica, Universidade Federal de
Pernambuco, Av. Jorn. An\'ibal Fernandes, s/n, Cidade
Universit\'aria 50740-560, Recife, PE, Brazil} \affiliation{Programa
de P\'os-Gradua\c{c}\~{a}o em Ci\^encia de Materiais, Universidade
Federal de Pernambuco, Av. Jorn. An\'ibal Fernandes, s/n, Cidade
Universitaria 50740-560, Recife, PE, Brazil}

\date{today}

\begin{abstract}
It is well-known that the appearance of almost-empty (shallow)
conduction bands in solids strongly affects their superconducting
properties. In a shallow band charge carriers are depleted and have
nearly zero velocities so that the crossover from the
Bardeen-Cooper-Schrieffer (BCS) superfluidity to Bose-Einstein
condensation (BEC) is approached. Based on a two-band prototype
system with one shallow and one deep band, we demonstrate that the
fundamental phase diagram of the superconducting magnetic response
changes qualitatively as compared to standard superconductors with
only deep bands. The so-called intertype (IT) domain between
superconductivity types I and II systematically expands in the phase
diagram when passing from the BCS to BEC side: its width is
inversely proportional to the squared Cooper-pair radius that
shrinks several orders of magnitude through the crossover. We also
show that the coupling to a stable condensate of the deep band makes
the system rather robust against the otherwise strong
superconducting fluctuations. Thus, the BCS-BEC crossover induced by
a shallow band pushes standard superconductivity types wide apart so
that the IT domain tends to dominate the phase diagram and therefore
the magnetic properties of shallow-band superconductors.
\end{abstract}

\maketitle

The crossover from the Bardeen-Cooper-Schrieffer (BCS) superfluidity
to Bose-Einstein condensation (BEC) is usually investigated in
trapped ultracold fermionic gases with the resonant
scattering~\cite{zwi,regal,grimm,dal}. However, it was originally
proposed for solids with a shallow conduction band whose lower edge
is close to the chemical potential~\cite{eagl}, see also the review
in \cite{chen}. Although theoretical studies of the BCS-BEC
crossover in superconductors have a long history, its unambiguous
experimental evidences have been obtained only recently in ${\rm Fe
Se}_x{\rm Te}_{1-x}$, where shallow Fermi pockets were proved to
play a significant role~\cite{kan,kas,okaz}. Interest in such
superconducting materials is fueled by expectations of a higher
critical temperature $T_c$~\cite{bor,boz} and novel
multigap/multicondensate coherent phenomena~\cite{pm} potentially
useful for technological applications. Here we demonstrate that the
BCS-BEC crossover regime realized in a multiband superconductor can
profoundly influence the superconducting magnetic properties so that
the relevant phase diagram differs strikingly from the standard one.

Earlier investigations of the BCS-BEC crossover in a charged
superfluid considered a sharp interchange between types I and II
throughout the entire crossover interval~\cite{melo}. This
consideration was based on the Ginzburg-Landau (GL) theory according
to which the types I and II interchange when the GL parameter
$\kappa=\lambda /\xi$ ($\lambda$ and $\xi$ are the magnetic and
coherence lengths) crosses the critical value $\kappa_0 =
1/\sqrt{2}$~\cite{degen,landau9,kett}. However, it is known that the
results of the GL theory for the phase diagram of the
superconducting magnetic properties are valid only in the limit $T
\to T_c$. In particular, below $T_c$ the intertype (IT) regime is
not reduced to the single point $\kappa=\kappa_0$ but occupies a
finite temperature-dependent interval of $\kappa$'s, forming the IT
domain in the $(\kappa,T)$ plane~\cite{kumpf,jacobs1,jacobs2,
jacobs3, jacobs4,hubert,auer,weber,luk,extGL1}. Magnetic properties
of a superconductor in this domain are governed by the Bogomolnyi
duality between the magnetic field and the condensate
density~(self-duality)~\cite{bogomol1,bogomol2,luk,extGL1}, that
facilitates formation of exotic flux/condensate configurations such
as, e.g., a lattice of superconducting islands, stripe/labyrinths
patterns, giant vortices, and vortex clusters~\cite{cor}. In
conventional single-band superconductors the IT domain is almost
negligible and thus ignored in textbooks. However, in multiband
superconducting materials it shows a general tendency to increase
due to enhancement of the non-local effects~\cite{extGL1}.

This work demonstrates that the proximity to the BCS-BEC crossover
induced by a shallow band has a dramatic effect on the IT domain:
its width is inversely proportional to the squared Cooper-pair
radius which shrinks several orders of magnitude when passing from
the BCS to BEC regime~\cite{pist,shan1,shan2}. Our analysis is done
for a two-band prototype system with one shallow and one deep band,
where closed analytical results can be derived. However, a universal
character of the employed formalism allows one to expect
qualitatively similar results for systems with an arbitrary number
of bands. Our conclusions are obtained within the mean-field theory,
which remains valid despite the presence of a shallow band: coupling
to the stable condensate of the deep band screens the otherwise
strong superconducting fluctuations~\cite{per_var}.

The IT domain is described by the critical GL parameters
$\kappa_i^{\ast}$, each marking the appearance of a particular flux
configuration $i$ at the thermodynamic critical field $H_c$: at
$\kappa > \kappa_i^{\ast}$ configuration $i$ becomes more favorable
energetically than the Meissner state~\cite{jacobs1,jacobs2,jacobs3,
jacobs4,luk,extGL1}. The critical GL parameters are found from the
equation
\begin{align}
\mathfrak{G}(\kappa_i^{\ast},T) = 0, \quad \mathfrak{G} =\int d{\bf
r}\,\left(\mathfrak{f} +\frac{H^2_c}{8\pi}- \frac{H_c
B}{4\pi}\right), \label{eq:critical_parameter}
\end{align}
where $\mathfrak{G}$ is the difference between the Gibbs free-energy
of configuration $i$ and of the Meissner state at $H_c$, and
$\mathfrak{f}$ is the condensate free-energy density. The magnetic
induction ${\bf B}$ is assumed to be parallel to the external field
${\bf H}$. Equation~(\ref{eq:critical_parameter}) can also be used
to find a critical GL parameter associated with qualitative changes
in some properties of a mixed state, e.g., in the sign of the
long-range vortex-vortex interaction~\cite{jacobs4,hubert}. Due to
the Bogomolnyi self-duality, the GL theory is infinitely degenerate
at $\kappa=\kappa_0$ and $H=H_c$~\cite{bogomol1,bogomol2,luk,
extGL1}. Then it predicts that all critical GL parameters are equal
to $\kappa_0$~\cite{luk,extGL1,weinb}. However, when corrections to
the GL theory are taken into account, the degeneracy is removed,
giving rise to an infinite number of different temperature dependent
$\kappa^\ast_i$, that shape the IT domain in the $(\kappa,T)$ plane.
Its lower boundary $\kappa^\ast_{min}$ is found from the onset of
the superconductivity nucleation at $H_c$, which is equivalent to
the condition $H_{c2} = H_c$, with $H_{c2}$ the upper critical
field. The upper boundary $\kappa^\ast_{max}$ is determined by the
appearance of the long-range attractive Abrikosov vortices. At $T
\to T_c$ the GL theory is exact and thus in this limit
$\kappa^\ast_i \to \kappa_0$ for all $i$.

We calculate $\kappa_i^\ast$ using the extended GL (EGL) formalism
that incorporates the leading corrections to the GL theory within
the perturbation expansion of the BCS equations over $\tau=1 -
T/T_c$~\cite{extGL2,extGL3,extGL4}. Here a sketch of the derivation
is presented where we highlight differences with the earlier
works~\cite{extGL1,extGL3} that appear due to the presence of the
shallow band. The BCS free-energy density for a two-band system
writes as
\begin{align}
\mathfrak{f}=\frac{{\bf B}^2}{8\pi} +\Delta^{\dagger}
\check{g}^{-1}\Delta + \sum\limits_{\nu=1,2}\mathfrak{f}_{\nu},
\label{eq:f}
\end{align}
where the vector $\Delta^{\dagger}=\big(\Delta^\ast_1,\Delta^\ast_2
\big)$ comprises the band gap functions $\Delta_\nu$,
$\check{g}^{-1}$ is the inverse of the coupling matrix $\check{g}$
($g_{ij} = g_{ji}$ are real) and $\mathfrak{f}_{\nu}$ is a
functional of $\Delta_\nu$. Expanding $\mathfrak{f}_{\nu}$ in powers
of $\Delta_\nu$ and its gradients, one finds~\cite{extGL1,extGL3}
\begin{align}
\mathfrak{f}_{\nu} = &- a_{1,\nu} |\Delta_\nu|^2 + a_{2,\nu} |{\bf
D}\Delta_\nu|^2 - a_{3,\nu}\Big(|{\bf D}^2
\Delta_\nu|^2 \notag \\
& +\frac{{\rm rot}{\bf B}\cdot{\bf i}_{\nu}}{3}+ \frac{4e^2}{\hbar^2
\mathbbm{c}^2}{\bf B}^2 |\Delta_\nu|^2\Big)
+ a_{4,\nu}{\bf B}^2|\Delta_\nu|^2 \notag \\
& +\frac{b_{1,\nu}}{2}|\Delta_\nu|^4 -\frac{b_{2,\nu}}{2}
\Big(L_{\nu} |\Delta_\nu|^2|{\bf D}\Delta_\nu|^2 \notag\\
&+l_{\nu} \big[(\Delta_\nu^{\ast})^2 ({\bf D}\Delta_\nu)^2 + {\rm
c.c.}\big]\Big)- \frac{c_{1,\nu}}{3}|\Delta_\nu|^6, \label{eq:f_nu}
\end{align}
where $a_{n,\nu}$, $b_{n,\nu}$, $c_{n,\nu}$ are
temperature-dependent band coefficients, $L_{\nu}$ and $l_{\nu}$ are
constants introduced here to capture qualitative differences (e.g.,
dimensionality, depth, etc.) between the contributing bands (compare
Eq.~(\ref{eq:f_nu}) with Eq.~(27) in \cite{extGL3}) and
\begin{align}
{\bf i}_{\nu} = -\frac{4 e}{\hbar\,\mathbbm{c}}{\rm Im} \big[
\Delta_{\nu}{\bf D}^\ast \,\Delta_{\nu}^\ast \big], \quad
\boldsymbol{D}=\boldsymbol{\nabla}- \mathbbm{i}\frac{2e}
{\hbar\,\mathbbm{c}} \boldsymbol{A}. \label{eq:i_nu}
\end{align}
The $\tau$-expansion is obtained by representing all entering
quantities as $\tau$-series
\begin{align}
&\Delta =\tau^{1/2} \big[\Delta^{(0)} + \tau \Delta^{(1)}\big],\;
{\bf B} = \tau \big[{\bf B}^{(0)} + \tau {\bf B}^{(1)}\big],\notag\\
&{\bf A} = \tau^{1/2} \big[{\bf A}^{(0)} + \tau {\bf
A}^{(1)}\big],\; H_c = \tau \big[H^{(0)}_c +\tau H^{(1)}_c\big],
\label{eq:DelBA}
\end{align}
where the lowest-order (GL) contributions and leading corrections
are given. We also invoke the $\tau$-scaling of the
coordinates~\cite{extGL3,extGL4} which leads to the additional
factor $\tau^{1/2}$ for each gradient in the $\tau$-expansion of the
functional. The temperature-dependent band coefficients in
Eq.~(\ref{eq:f_nu}) are obtained as
\begin{align}
&a_{1,\nu} = {\cal A}_{\nu} - \tau\big[a^{(0)}_{\nu} + \tau
a^{(1)}_{\nu}\big],\; a_{2,\nu} = {\cal K}^{(0)}_{\nu} +
\tau{\cal K}^{(1)}_{\nu},\notag\\
&a_{3,\nu}={\cal Q}^{(0)}_{\nu},\quad a_{4,\nu} = r^{(0)}_{\nu},
\quad b_{1,\nu} = b^{(0)}_{\nu}+\tau b^{(1)}_{\nu},\notag \\
&b_{2,\nu}L_{\nu}= {\cal L}^{(0)}_{\nu},\quad b_{2,\nu}l_{\nu}=
\ell^{(0)}_{\nu},\quad c_{1,\nu}= c^{(0)}_{\nu}, \label{eq:coef_exp}
\end{align}
where the $\tau$-expansion coefficients are calculated using a
particular microscopic model of the band. Substituting
Eqs.~(\ref{eq:DelBA}), (\ref{eq:coef_exp}) and the gradient scaling
into Eqs.~(\ref{eq:f}) and (\ref{eq:f_nu}), one obtains the
$\tau$-expansion for the free-energy density. It is then inserted in
Eq.~(\ref{eq:critical_parameter}), which gives the corresponding
series for $\mathfrak{G}$.

\begin{table*}
\begin{ruledtabular}
\begin{tabular}{ccccccccc}
{\color{black} $\nu$} & ${\cal M}^{(0)}_{b,\nu}$ & ${\cal
M}^{(0)}_{c,\nu}$ & ${\cal M}^{(0)}_{{\cal K},\nu}$ & ${\cal
M}^{(0)}_{{\cal Q},\nu}$ & ${\cal M}^{(0)}_{{\cal L},\nu}$ & ${\cal
M}^{(1)}_{a,\nu}$ & ${\cal M}^{(1)}_{b,\nu}$ & ${\cal
M}^{(1)}_{{\cal K},\nu}$\\[1ex]
\hline
\\
1 & $7\zeta(3)/(8\pi^2)$ & $93\zeta(5)/(128\pi^4)$ &
$7\zeta(3)/(32\pi^2)$ & $93\zeta(5)/(2048\pi^4)$ &
$31\zeta(5)/(32\pi^4)$ & 1/2 & 2 & 2\\[1ex]
2 & $7\zeta(3)/(8\pi^2)$ & $93\zeta(5)/(128\pi^4)$ &
$3\zeta(2)/(8\pi^2)$  & $7\zeta(3)/(512\pi^2)$ &
$25\zeta(4)/(16\pi^4)$ & 1/2
& 2 & 1\\[1ex]
\end{tabular}
\end{ruledtabular}
\label{tab:tab1} \caption{Numerical factors ${\cal
M}^{(0)}_{w,\nu}$~(with $w=b,c,{\cal K},{\cal Q},{\cal K}$) and
${\cal M}^{(1)}_{w,\nu}$~(with $w=a,b,{\cal K}$) for the deep
($\nu=1$) and shallow ($\nu=2$) bands, see Eqs.~(\ref{eq:c_kappa4A})
and (\ref{eq:c_kappa4B}). In the table $\zeta(x)$ is the Riemann
zeta function of $x$.}
\end{table*}

To obtain the two lowest orders in the series for $\mathfrak{G}$,
one needs only $\Delta_{1,2}^{(0)}$ and ${\bf B}^{(0)}$ because the
contributions containing $\Delta_{\nu}^{(1)}$ and ${\bf B}^{(1)}$
are either zero (up to the vanishing surface integrals) or can be
rewritten through $\Delta_{1,2}^{(0)}$ and ${\bf
B}^{(0)}$~\cite{extGL1}. As a result, the two leading contributions
to $\mathfrak{G}$ can be calculated using only the solution to the
GL formalism. The latter takes the form
\begin{align}
 \left(
\begin{array}{c}
\Delta_1^{(0)}\\
\Delta_2^{(0)}
\end{array}
\right) = \Psi({\bf r})\,\left(
\begin{array}{c}
S^{-1/2}\\
S^{1/2}
\end{array}
\right), \label{eq:Psi}
\end{align}
where $\Psi$ is the Landau order parameter that satisfies the
single-component GL equation, and $S$ is determined by the
linearized gap equation that yields
\begin{align}
S = \frac{1}{g_{12}} \big(g_{22} - G {\cal A}_1\big)=
\frac{g_{12}}{g_{11}-G{\cal A}_2},
\label{eq:S}
\end{align}
with $G={\rm det}\,\check{g}=g_{11}g_{22}-g_{12}^2$.

The critical GL parameters are sought in the form $\kappa^{\ast} =
\kappa_0 + \delta \kappa$, where $\delta \kappa \sim \tau$~(we hide
the index $i$ unless it causes confusion). Then, we utilize a
self-dual form of the GL theory (the self-duality Bogomolnyi
equations)~\cite{bogomol1,bogomol2} which allows one to explicitly
express ${\bf B}^{(0)}$ as a function of $|\Psi|^2$. The resulting
expansion of $\mathfrak{G}$ contains only the linear terms $\propto
\tau$ and $\propto \delta \kappa$, while the GL contribution
vanishes due to the degeneracy of the GL theory at $\kappa_0$ and
$H_c$. Resolving Eq.~(\ref{eq:critical_parameter}), one obtains
$\kappa^\ast = \kappa_0 + \tau \kappa^{\ast(1)}$ with
\begin{align}
\frac{\kappa^{\ast(1)}}{\kappa_0} =&\;\bar{\cal K} -\bar{c}
+2\bar{\cal Q}+\bar{G}\,\bar{\beta}\big(2\bar{\alpha}
-\bar{\beta}\big)\notag\\
&+\frac{{\cal J}}{\cal I}\left(\frac{\bar{\cal L}}{4} - \bar{c}
-\frac{5}{3}\bar{\cal Q} - \bar{G}\bar{\beta}^2\right),
\label{eq:kappa1}
\end{align}
where the dependence on a particular mixed-state configuration
enters via the integrals
\begin{align}
{\cal I} =\!\! \int |\Psi|^2 \big(1 - |\Psi|^2\big)d{\bf
r},\;\,\,{\cal J} =\!\! \int |\Psi|^4 \big(1 - |\Psi|^2\big)d{\bf
r}. \label{eq:I_J}
\end{align}
Dimensionless constants in Eq.~(\ref{eq:kappa1}) are given by
\begin{align}
&\bar{\cal K} = \frac{{\cal K}^{(1)}}{\cal K}-\frac{{b}^{(1)}}{2b},
\; \bar c =\frac{c a }{3 b^2},\; \bar {\cal Q}=\frac{a{\cal
Q}}{{\cal K}^2},\; \bar{\cal L}=\frac{a {\cal L}}{b{\cal K}},
\notag\\
&\bar{G} =\frac{G a}{4g_{12}}, \quad \bar
\alpha=\frac{\alpha}{a}-\frac{{\Gamma}}{{\cal K}}, \quad  \bar\beta
= \frac{\beta}{b}-\frac{{\Gamma}}{{\cal K}}, \label{eq:c_kappa1}
\end{align}
where
\begin{align}
&w = w_1^{(0)}S^{-p} +S^p w_2^{(0)},\;
w^{(1)}=w_1^{(1)}S^{-p} + S^p w_2^{(1)},  \notag \\
&\alpha=a_1^{(0)}S^{-1} - S a_2^{(0)},\;
\beta=b_1^{(0)}S^{-2}-S^2 b_2^{(0)},\notag \\
&\Gamma = {\cal K}_1^{(0)}S^{-1}-S{\cal K}_2^{(0)}.
\label{eq:c_kappa2}
\end{align}
with the substitutions $w = \{a,{\cal K},{\cal Q},b,{\cal L},c\}$,
$w_\nu^{(0)} = \{a_\nu^{(0)},{\cal K}_\nu^{(0)},{\cal Q}_\nu^{(0)},
b_\nu^{(0)},{\cal L}_\nu^{(0)},c_\nu^{(0)}\}$, $w^{(1)} = \{ {\cal
K}^{(1)}, b^{(1)}\}$, and $w_\nu^{(1)} = \{{\cal K}_\nu^{(1)},
b_\nu^{(1)}\}$. In Eq.~(\ref{eq:c_kappa2}) $p=1$ appears in $w$'s
related to $a_{n,\nu}$ while $p=2$ and $3$ correspond to
$b_{n,\nu}$, and $c_{n,\nu}$, respectively. Notice that the terms
containing $a^{(1)}_{\nu}$ and $\ell^{(0)}_{\nu}$ do not contribute
to $\mathfrak{G}$ and, thus, to Eq.~(\ref{eq:kappa1}). In turn,
$r_\nu^{(0)}$ is negligible and so ignored. One notes that
Eqs.~(\ref{eq:kappa1})-(\ref{eq:c_kappa2}) differ from the results
obtained for a system with two deep bands in Ref.~\cite{extGL1}:
here the expression for $\kappa^\ast$ contains additional terms.

Coefficients in Eq.~(\ref{eq:coef_exp}) are calculated assuming that
both bands have a 2D circular-symmetry Fermi surface with the band
single-particle dispersion $\varepsilon_{\nu,k} =\varepsilon_{\nu,0}
+ (\hbar^2/2m_\nu) (k^2_x +k^2_y)$, where $\varepsilon_{\nu,0}$ is
the band lower edge and $m_\nu$ is the band carrier mass. The
magnetic field is chosen in the $z$ direction to deal with the
isotropic system. The calculations are performed in the clean limit.
For the deep band ($\nu=1$), one employs the standard BCS
approximations~\cite{degen,landau9,kett} as $\Delta_1 \ll \mu -
\varepsilon_{1,0}$. Calculations for the shallow band ($\nu = 2$)
are technically more involved. However, analytic expressions for all
coefficients can be derived when the chemical potential touches the
band lower edge, i.e., $\mu = \varepsilon_{2,0}$~(this can be
assumed without the generality loss). Notice, that the 2D character
of the shallow band is important: although the band is almost empty,
its DOS at the lower edge remains sizeable. Calculations yield the
following lowest-order coefficients in Eq.~(\ref{eq:coef_exp}):
\begin{align}
& {\cal A}_\nu = N_\nu \ln\Big(\frac{2e^\gamma\hbar \omega_c}{\pi
T_c}\Big),\, a^{(0)}_{\nu} = - N_{\nu},\,b^{(0)}_{\nu} = N_{\nu}
\frac{{\cal M}^{(0)}_{b,\nu}}{T_c^2},\notag\\
&c^{(0)}_{\nu} =N_{\nu}\frac{{\cal M}^{(0)}_{c,\nu}}{T_c^4},\, {\cal
K}^{(0)}_{\nu} = N_{\nu}{\cal M}^{(0)}_{\cal
K,\nu}\frac{\hbar^2v_{\nu}^2}{T_c^2},\notag\\
&{\cal Q}^{(0)}_{\nu} = N_{\nu} {\cal M}^{(0)}_{{\cal Q},{\nu}}
\frac{\hbar^4 v_{{\nu}}^4}{T_c^4},\;{\cal L}^{(0)}_{\nu} =
N_{\nu}\,{\cal M}^{(0)}_{{\cal L},\nu}\frac{\hbar^2
v_{\nu}^2}{T_c^4},
\label{eq:c_kappa4A}
\end{align}
where $\hbar\omega_c$ is the cut-off energy, $\gamma$ is the Euler
constant, $N_{\nu}$ is the band DOS, and $v_{\nu}$ denotes the
characteristic band velocity, i.e., the Fermi velocity $v_F =
\sqrt{2(\mu-\varepsilon_{1,0})/m_1}$ for the deep band and the
temperature-driven velocity $v_T =\sqrt{2T_c/m_2}$ for the shallow
band. The factors ${\cal M}^{(0)}_{w,\nu}$ are given in Tab. I. The
band DOS are $N_1 = \tilde{N}_1 m_1/(2\pi\hbar^2)$ and
$N_2=\tilde{N}_2 m_2/(4\pi\hbar^2)$, with $\tilde{N}_\nu$ introduced
to account for the states in the $z$ direction (this quantity does
not affect final conclusions). The next-order coefficients in
Eq.~(\ref{eq:coef_exp}) are obtained as
\begin{align}
w^{(1)}_{\nu}={\cal M}^{(1)}_{w,\nu}w^{(0)}_{\nu},
\label{eq:c_kappa4B}
\end{align}
where $w=\{a,{\cal K},b\}$ and ${\cal M}^{(1)}_{w,\nu}$ are also
shown in Tab. I.

\begin{figure}[t]
\begin{center}
\resizebox{1.0\columnwidth}{!}{\rotatebox{0}{
\includegraphics{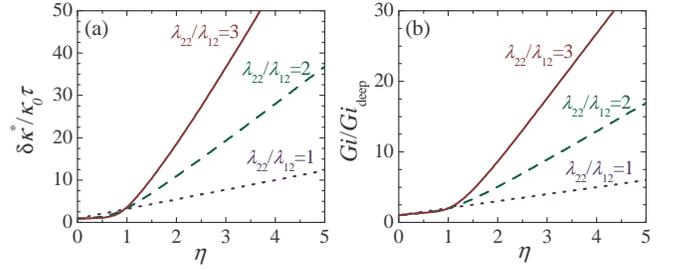}}}
\end{center}
\caption{(a) The size of the intertype domain (relative to
$\kappa_0\tau$) as a function of $\eta$ calculated at $v_2/v_1 = 0$
and $\lambda_{11}=\lambda_{22} =0.3$ for $\lambda_{22}/\lambda_{12}
=1$~(dotted), $2$~(dashed), and $3$~(solid). (b) The relative
Ginzburg number $Gi/Gi_{\rm deep}$ versus $\eta$ obtained for the
same parameters as in panel (a).} \label{fig1}
\end{figure}

We can now calculate the boundaries of the IT domain.
$\kappa^\ast_{min}$ is obtained from the condition that the
inhomogeneous mixed state disappears at $H_c$, which gives ${\cal
J}/{\cal I} = 0$~(as $\Psi \to 0$). At $\kappa=\kappa^\ast_{max}$ he
long-range vortex-vortex interaction changes its sign. For the
two-vortex solution one finds ${\cal J}(R)/{\cal I}(R) \to 2$ in the
limit of large inter-vortex distance $R \to \infty$~\cite{extGL1}.
Using these results, one obtains final expressions for the critical
GL parameters $\kappa^\ast_{min}$ and $\kappa^\ast_{max}$ as
complicated algebraic functions of the microscopic parameters $\eta
= N_2/N_1$, $v_2/v_1$, and $g_{ij}$. Obtained expressions, however,
can be simplified because $v_2/v_1 \sim \sqrt{T_c/\mu} \ll 1$ and
$\eta \gg 1$~(the latter inequality is dictated by the proximity to
the BCS-BEC crossover \cite{shan1, shan2}). In this case the
difference $\delta \kappa^\ast = \kappa^{\ast}_{\rm
max}-\kappa^\ast_{\rm min}$ reduces to
\begin{align}
\frac{\delta \kappa^\ast }{\kappa_0\tau}
\simeq -\frac{10}{3} \bar{\cal Q} \simeq 2.27
\left(\frac{\lambda_{22}}{\lambda_{12}}\right)^2\!\!\eta,
\label{eq:domain}
\end{align}
where $\bar{\cal Q} \simeq [93\zeta(5)/98\zeta^2(3)] S^2 \eta$ and
$S$ approaches $\lambda_{22}/\lambda_{12}$, with the dimensionless
coupling $\lambda_{ij}=g_{ij}N$~($N=N_1+N_2$). The expression for
$S$ follows from the solution to Eq.~(\ref{eq:S}), i.e.,
\begin{align}
S=\frac{1}{2\lambda_{12}}\!\left[\lambda_{22}-
\frac{\lambda_{11}}{\eta}
+ \sqrt{\Big(\lambda_{22}-\frac{\lambda_{11}}{\eta}\Big)^2\!\!
+4\frac{\lambda^2_{12}}{\eta}}\right]. \label{eq:S_eta}
\end{align}
Equation~(\ref{eq:domain}) demonstrates that the IT domain
systematically enlarges when approaching the BCS-BEC crossover. This
enlargement is more pronounced when increasing
$\lambda_{22}/\lambda_{12}$, i.e., when the role of the deep band is
further diminished. It is important to note that the dominant
contribution to the enlargement given by Eq.~(\ref{eq:domain}) is
provided by the most non-local terms in the free energy functional,
i.e., those with the fourth order gradients, see
Eq.~(\ref{eq:f_nu}). This is closely connected to the known fact
that the IT domain appears due to non-local interactions in the
condensate beyond the standard GL theory~\cite{rev_brandt}.
Numerical results for $\delta\kappa^\ast$ given in
Fig.~\ref{fig1}(a) as a function of $\eta$, are calculated using the
original not simplified expressions at $v_2/v_1=0$, where we set
$\lambda_{11} =\lambda_{22}= 0.3$ and $\lambda_{22}/ \lambda_{12}=
1,2$ and $3$~(results are qualitatively similar for any choice of
the coupling constants). One can see that $\delta\kappa^\ast$
exhibits a linear dependence of the simplified Eq.~(\ref{eq:domain})
already at moderate values of $\eta \gtrsim 1$.

Finally we address the role of the superconducting fluctuations. It
is commonly expected that the fluctuations are enhanced in
superconductors with shallow bands, which can compromise the
mean-field results. However, in multiband systems even a weak
coupling to a deep band can screen the fluctuations~\cite{per_var}.
We estimate their relative importance by calculating the Ginzburg
number $Gi$, i.e., the temperature interval near $T_c$ where the
fluctuation-induced contribution to the heat capacity $\delta c_V $
exceeds its mean-field counterpart $c_{0,V}$~\cite{kett}. For bands
with 2D Fermi surfaces we get
\begin{align}
c_{0,V} = \frac{a^2}{bT_c},\quad \delta c_V \simeq \frac{1}{4\pi
\xi^2_0d_z\tau},
\label{eq:CV}
\end{align}
where $a,b$ are given by Eq.~(\ref{eq:c_kappa2}), and $d_z/2\pi$ is
the inverse of the size of the Brillouine zone in the $z$ direction.
The Cooper-pair radius (the BCS coherence length) $\xi_0$ is found
for the two-band system as
\begin{align}
\xi^2_0 = - \frac{{\cal K}}{a} =  \frac{\xi^2_{0,1}}{1 + \eta S^2} +
\frac{\xi^2_{0,2}}{1 + \eta^{-1} S^{-2}}, \label{eq:xi20}
\end{align}
where $\xi_{0,\nu}^2= -{\cal K}^{(0)}_\nu/a^{(0)}_\nu$ are the band
BCS lengths. Taking into account that $\xi_0$ approaches
$\xi_{2,0}\approx 0$ at $\eta \to \infty$, one can see that in this
limit $\delta c_V$ sharply increases. Still, unless $\eta$ assumes
extreme values, the contribution of the deep band to $\xi_0$ remains
dominant even at very large $\eta$ as long as $\xi_{0,1} \gg
\xi_{0,2}$. As a result $\xi^2_0$ and $\delta c_V$ remain close to
their deep-band values, which is seen as the screening of the
superconducting fluctuations~\cite{per_var}. $Gi$ is obtained by
resolving $c_{0,V} = \delta c_V|_{\tau=Gi}$,
\begin{align}
Gi = Gi_{\rm deep}\frac{(1+\eta)(1+\eta S^4)}{1+\eta S^2},
\label{eq:Gi}
\end{align}
where $Gi_{\rm deep}$ is calculated for the deep band with the DOS
$N = N_1 + N_2$. For $v_2/v_1 =0$ and $\eta \to \infty$,
Eq.~(\ref{eq:Gi}) yields $Gi/Gi_{\rm deep} \simeq \eta
(\lambda_{22}/\lambda_{12})^2$. Comparing Eqs.~(\ref{eq:domain}),
(\ref{eq:xi20}) and (\ref{eq:Gi}), we arrive at
\begin{align}
\delta\kappa^\ast/\kappa_0 \propto Gi/Gi_{\rm deep} \propto
(k_F\xi_0)^{-2},
\end{align}
which relates the width of the IT domain and the $Gi$ parameter with
the Cooper-pair radius given in units of $k_F^{-1}$~($k_F$ is the
Fermi wavenumber in the deep band). We note that it is a comparison
of $\xi_0$ with the average interparticle distance that defines the
proximity to the BCS-BEC crossover, and in the case of interest this
distance is of the order of $k_F^{-1}$. Taking into account that for
typical superconducting systems $Gi_{\rm deep} \sim 10^{-16}\div
10^{-6}$~\cite{kett}, one concludes that the presence of a shallow
band does not invalidate the mean-field approach in a very wide
parameter range: the fluctuations can be neglected even if $\xi_0$
drops by several orders of magnitude. For illustration
Fig.~\ref{fig1}(b) shows $Gi/Gi_{\rm deep}$ as a function of $\eta$,
calculated for the same parameters as the results for $\delta
\kappa^\ast$ in panel (a). One sees that, similarly to
$\delta\kappa^\ast$, $Gi$ quickly approaches the linear asymptotic
regime at $\eta \gtrsim 1$.

In summary, we have demonstrated that the BCS-BEC crossover, induced
by the presence of a shallow band, pushes apart standard
superconductivity types I and II by enhancing the IT domain of
unconventional superconducting magnetic properties. This effect is
mainly controlled by the highest-gradient contribution to the free
energy functional. Although our analysis focuses on the two-band
model, we expect that our conclusions hold for a general multiband
superconductor with at least one shallow and one deep band in the
electronic spectrum. The reason is that the EGL formalism yields
formally similar results for an arbitrary number of bands, as long
as there is no additional symmetry in the system, see \cite{extGL3}.
We note that our choice of the dimensionality for the deep band is
not crucial and qualitatively similar results can be obtained for
the deep band with a 3D Fermi surface. On the contrary, when
assuming a 3D Fermi surface for the shallow band, the BCS-BEC
crossover can be reached only for an abnormally strong coupling due
to negligibly small DOS~\cite{gud}. Finally, the calculations of the
Ginzburg number have confirmed that the presence of a shallow
band(s) in multiband superconductors does not compromise the
mean-field approach in a wide parameter range because of the
screening effects generated by the coupling to deep bands.

The authors acknowledge support from the Brazilian agencies CNPq
(grants 307552/2012-8 and 141911/2012-3) and FACEPE
(APQ-0936-1.05/15).

\end{document}